%% file: main.tex
\documentclass[letterpaper]{article} 
\usepackage{aaai2026}  
\usepackage{times}  
\usepackage{helvet}  
\usepackage{courier}  
\usepackage[hyphens]{url}  
\usepackage{graphicx} 
\usepackage{natbib}  
\usepackage{caption} 
\usepackage{algorithm}
\usepackage{algorithmic}
\usepackage{amssymb}
\usepackage{amsmath}
\usepackage{amsthm}
\usepackage{bbding}
\usepackage{booktabs}
\usepackage{multirow}
\usepackage{subfigure}
\usepackage{colortbl}
\usepackage{tcolorbox}

\usepackage{newfloat}
\usepackage{listings}
\DeclareCaptionStyle{ruled}{labelfont=normalfont,labelsep=colon,strut=off} 
\floatstyle{ruled}
\newfloat{listing}{tb}{lst}{}
\floatname{listing}{Listing}

\title{M\textsuperscript{2}VAE: Multi-Modal Multi-View Variational Autoencoder for Cold-start Item Recommendation}
\author{
    Chuan He\textsuperscript{\rm 1,2}\thanks{This work was done during a research internship at Ant Group.},
    Yongchao Liu\textsuperscript{\rm 2\dag},
    Qiang Li\textsuperscript{\rm 3},
    Chuntao Hong\textsuperscript{\rm 2},
    Wenliang Zhong\textsuperscript{\rm 2},
    Xin-Wei Yao\textsuperscript{\rm 3}\thanks{Co-corresponding authors.}
}
\affiliations{
    \textsuperscript{\rm 1}College of Computer Science and Technology, Zhejiang University of Technology, Hangzhou, China\\
    \textsuperscript{\rm 2}Ant Group, Hangzhou, China\\
    \textsuperscript{\rm 3}Academy for Advanced Interdisciplinary Science and Technology, Zhejiang University of Technology, Hangzhou, China
    \{hechuan,qiangli,xwyao\}@zjut.edu.cn,\{yongchao.ly,chuntao.hct\}@antgroup.com,yice.zwl@alibaba-inc.com

}

\usepackage{bibentry}

\begin{document}
\maketitle

\input{section/abstract}

\input{section/intro}

\input{section/related_work}

\input{section/pre}  

\input{section/method}


\input{section/experiment}

\input{section/conclusion}

\input{section/ack}

\bibliography{reference}

\end{document}

%% file: section/abstract.tex
\begin{abstract}
 Cold-start item recommendation is a significant challenge in recommendation systems, particularly when new items are introduced without any historical interaction data. While existing methods leverage multi-modal content to alleviate the cold-start issue, they often neglect the inherent multi-view structure of modalities, namely the distinction between shared and modality-specific features. In this paper, we propose \underline{M}ulti-Modal \underline{M}ulti-View \underline{V}ariational \underline{A}uto\underline{E}ncoder (M\textsuperscript{2}VAE), a generative model that addresses the challenges of modeling common and unique views in attribute and multi-modal features, as well as user preferences over single-typed item features. Specifically, we generate type-specific latent variables for item IDs, categorical attributes, and image features, and use Product-of-Experts (PoE) to derive a common representation. A disentangled contrastive loss decouples the common view from unique views while preserving feature informativeness. To model user inclinations, we employ a user-aware hierarchical Mixture-of-Experts (MoE) to adaptively fuse representations. We further incorporate co-occurrence signals via contrastive learning, eliminating the need for pretraining. Extensive experiments on real-world datasets validate the effectiveness of our approach.
\end{abstract}

%% file: section/intro.tex
\section{Introduction}
As online information on e-Commerce and social media platforms continues to expand rapidly, recommender systems have become essential to help users navigate information overload. These systems facilitate the discovery of appealing products or content. However, when there is a scarcity of user-item interaction data, it becomes difficult to develop effective representations for users or items, resulting in the well-known cold-start problem.
\begin{figure}[h]
  \centering
  \includegraphics[width=\linewidth]{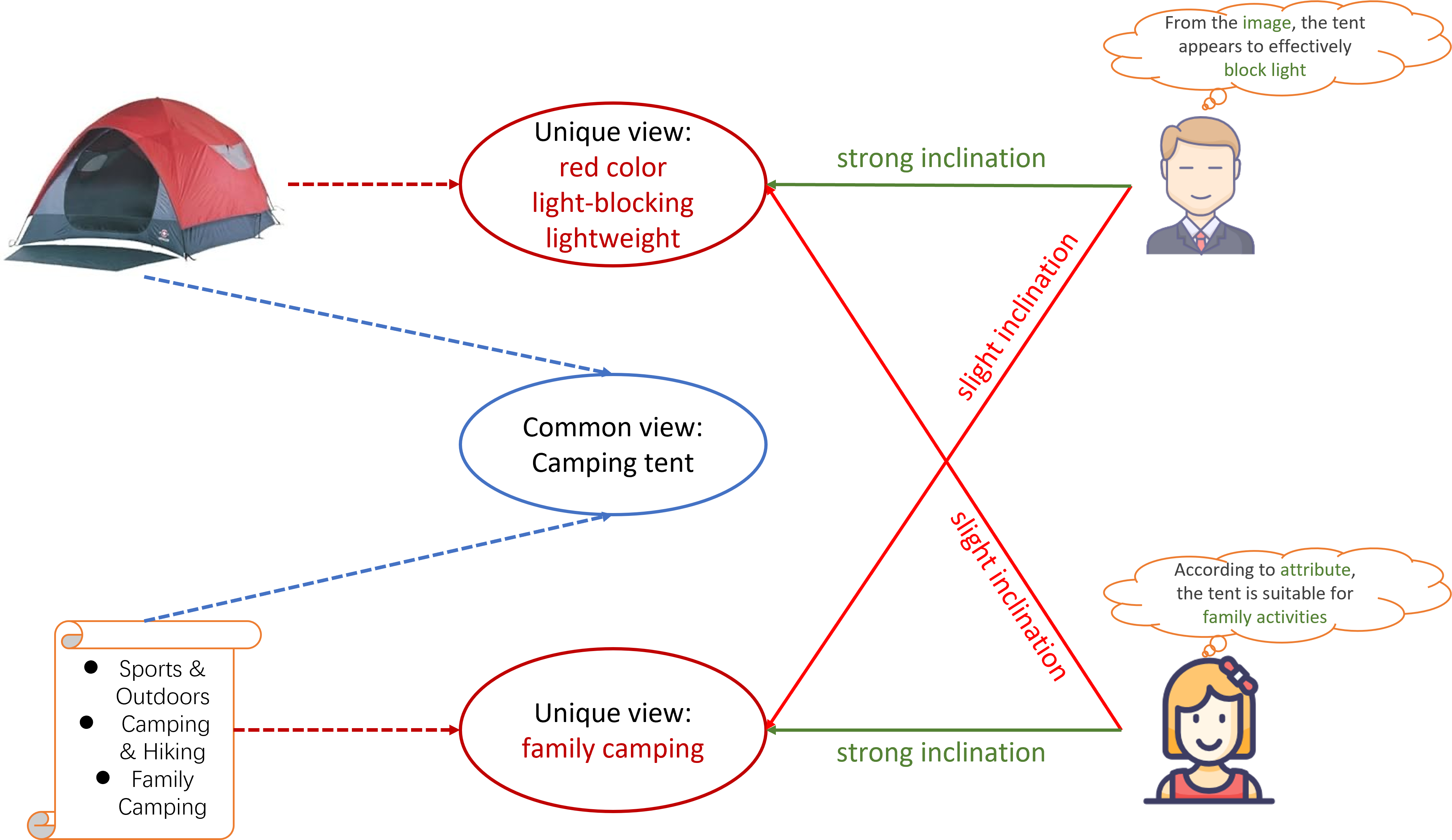}
  \caption{Observations in cold-start recommendation}
  \vspace{-0.3cm}
\end{figure}
A promising avenue of research focuses on content-based methods, which typically involve training a generative model to map the content of cold start items, such as attributes, text, images, and so on, into the embedding space of warm items \cite{GoRec, GME, GAR, Jcvae, JMVAE, MULTI-VAE}. For example, DropoutNet\cite{DropoutNet} facilitates this transformation by randomly omitting warm item embeddings learned during the training phase, allowing implicit conversion of cold start item content into a warm embedding. Meanwhile, MWUF\cite{MWUF} employs two meta networks to generate warm embeddings for cold-start items using feature and ID embeddings. Recently, variational auto-encoders (VAEs)\cite{VAE, RecVAE} have been widely utilized to tackle cold-start issues in recommendation. For example, GoRec\cite{GoRec} uses a CVAE model to reconstruct pre-trained preference representations and generate preference representations for new items.

Although generative methods(e.g., VAE-based, GAN-based) have demonstrated success in generating efficient representations for both items and users, they still face several challenges and limitations according to our observations:

\begin{itemize}
\item \textbf{Observation1. The presence of both common and unique perspectives within multi-typed item features.}  For example, consider a camping tent with categorical features such as ['Sports\&Outdoors', 'Camping\&Hiking', 'Family Camping Tents'] and an associated image feature shown in Figure 1. Both the categorical and image features collectively indicate that the item is a camping tent, reflecting the common view. However, when examining unique views, the categorical data specifies its suitability for family camping, while the image reveals details such as color, lightweight and light-blocking capabilities. This example underscores the need for multi-view feature encoding capabilities in VAE-based models to effectively integrate and leverage the diverse information provided by multi-typed features. Nevertheless, existing VAE-based methods, like GoRec\cite{GoRec} and MVGAE\cite{MVGAE}, encode the multi-typed item feature using simple Concat\&MLP in observation space or PoE fusion in latent space.
\item  \textbf{Observation2. Users' personal preferences toward the unique views of these features.} Some users may prioritize portability, focusing on the tent's lightweight design, others might value its light-blocking capabilities for a better sleep quality as shown in the image feature. Additionally, some users may rely on the category information to ensure the tent is suitable for their specific needs, such as family camping. Thus, modeling the personalized inclination to various unique views plays an important role in recommending a cold-start item. However, few methods explicitly model such user inclinations.
\end{itemize}

In this paper, we propose a novel generative model, the \underline{M}ulti-Modal \underline{M}ulti-View \underline{V}ariational \underline{A}uto\underline{e}ncoder (M\textsuperscript{2}VAE) framework, for new item recommendation. M\textsuperscript{2}VAE generates comprehensive representations of new items by explicitly modeling the common and unique aspects of multi-typed item features and incorporating personalized user preferences.
Our approach begins by generating type-specific latent variables for item ID embeddings, categorical attributes, and image features. We then use a Product-of-Experts (PoE) mechanism to derive a common representation that captures the shared information across these feature types. To disentangle the common view from the unique views of each feature type, we introduce a contrastive loss, which is supplemented by a reconstruction loss to ensure the generated representations accurately reflect the original features.
To incorporate personalized user preferences, we employ a Mixture-of-Experts (MoE) to fuse the common and unique view representations, thereby generating multifaceted feature variables that capture diverse item characteristics. We then utilize another naive MoE to integrate the ID variable with the feature variables, enabling the learning of a joint distribution over ID embeddings and multi-typed feature representations within the conditional variational autoencoder framework.
Finally, we enhance the model by incorporating co-occurrence signals through contrastive learning, which replaces the need for a pretraining module by learning from positive and negative item pairs selected for each user. To summarize, the contributions in this paper are listed as follows,
\begin{itemize}
    \item We introduce a mutli-modal multi-view variational framework that separately models the \textit{common view} (shared semantics across modalities) and \textit{unique views} (modality-specific characteristics).
    \item We design a disentangled contrastive loss that ensures both separation between views and preservation of original feature information.
    \item We propose an adaptive fusion mechanism via Mixture-of-Experts (MoE), which dynamically integrates item representations based on user inclinations.
    \item Extensive experiments on real-world datasets demonstrate the superiority of M\textsuperscript{2}VAE over state-of-the-art methods in cold-start recommendation settings. Our code is available Our code is available at: https://github.com/hchchchchchchc/M2VAE.
\end{itemize}


%% file: section/related_work.tex
\section{Related Work}

\subsection{Cold-start Recommendation}
The cold-start problem—arising from missing interaction history for new users or items—is commonly addressed via content-based and robust learning methods. Content-based approaches exploit side information to alleviate data sparsity~\cite{EmerG,KNN,DUIF,NFM2}, with generative models further bridging ID embeddings and content features~\cite{GME}. Meta-learning frameworks like MELU~\cite{MELU}, M2EU~\cite{M2EU}, and MWUF~\cite{MWUF} generate warm embeddings by leveraging user features. GAN-based models (e.g., VAE-AR~\cite{VAE-AR}, LARA~\cite{LARA}, GAR~\cite{GAR}) and dropout strategies such as DropoutNet~\cite{DropoutNet} enhance robustness under sparse interactions. Distribution alignment methods (EQUAL~\cite{EQUAL}, ALDI~\cite{ALDI}) and graph-based approaches (MVDGAE~\cite{MVDGAE}, IHGNN~\cite{IHGNN}, Frizen~\cite{Frizen}) also improve cold-start representation learning.

\subsection{VAE-based Methods in Recommendation}
Variational Autoencoders (VAEs) are widely used in recommendation for modeling user-item interaction distributions~\cite{surveyVAE,surveyVAERec,MAHGA,multi-vae-cv,TMC-VAE,NIPS2016variational}. Multi-VAE~\cite{MULTI-VAE} adopts a multinomial likelihood with neural parameterization, while RecVAE~\cite{RecVAE} improves inference via a composite prior. Extensions include bilateral VAEs (BiVAE~\cite{BiVAE}), graph-integrated models (CVGA~\cite{CVGA}), and conditional variants like CVAE~\cite{CVAE}, MD-CVAE~\cite{MD-CVAE}, and GAR~\cite{GAR}, which align cold/warm embeddings or incorporate side information. Recent works such as SEM-MacriM\textsuperscript{2}VAE~\cite{SEM-MacridVAE}, CVAR~\cite{CVAR}, GoRec~\cite{GoRec}, and HCVAE~\cite{HCVAE} further enhance representation disentanglement and multi-source fusion, demonstrating VAEs’ versatility in recommender systems.

%% file: section/pre.tex
\section{Problem Formulation}
 In the cold-start item problem, due to the limited interactions between users and items, generating effective enough item ID embedding depends on categorical attribute and multi-modal content features. Let $\mathcal{U}(\lvert \mathcal{U} \rvert=M)$ and $\mathcal{V}(\lvert \mathcal{V} \rvert=N)$ be a set of users and a set of items, respectively. Each item $v_i \in \mathcal{V}$ is associated with a categorical attribute set $\mathcal{A}$ consist of n attributes $\{a_1, ..., a_n\}$, where an attribute is represented as a one-hot vector called categorical attribute feature. Meanwhile, each items is also associated with multi-modal content feature, e.g., item image. Let $\mathcal{C} \in \mathbb{R}^{N\times d}$ be a set of image feature of all items represented by real valued vectors. Let $\mathcal{V}_u \in \mathcal{V}$ be the set of items that user $u \in \mathcal{U}$ interacted with, which contains co-occurrence signals between users and warm items. The ultimate goal is to infer the probability $\hat{y}_{uv}$ user $u$ preferring new item $v$:
 \begin{equation}
     \hat{y}_{uv} = \mathcal{P}(\mathcal{F}_u(u), \mathcal{F}_v(v, A_v, C_v, \mathcal{V}_u))
 \end{equation}
 Where $\mathcal{F}_u$ and $\mathcal{F}_v$ are the functions to generating item and user representations. In this paper, we aim at generating cold-start item representation based on disentangled multi-typed feature representation fused by a method that embodies personalized inclination to unique views.

%% file: section/method.tex
\section{Methodology}
In this paper, we propose a Multi-modal Multi-view Variational AutoEncoder (M\textsuperscript{2}VAE), which is comprised of three components: multi-view generator, multi-view fusion, and co-occurrence signal injection, as illustrated in Fig. 2. Then, we illustrate the optimization of our proposed model. Finally, we comprehensively compare the fusion architecture with existing VAE-based methods and provide a rigorous theoretical analysis to verify the advantages of  M\textsuperscript{2}VAE.
\begin{figure*}[htbp]
    \centering
    \includegraphics[width=1.0\linewidth]{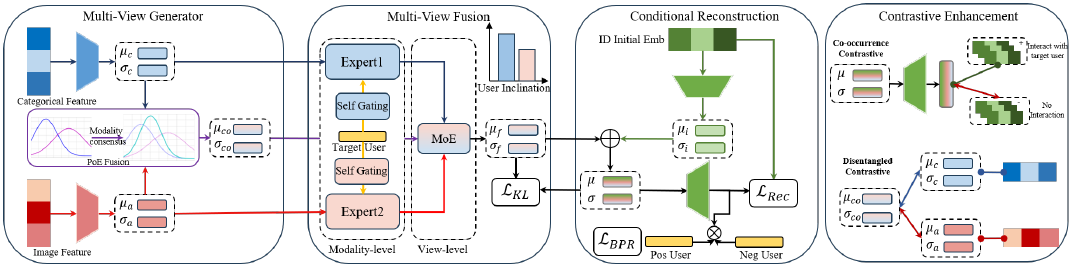}
    \caption{Framework of M\textsuperscript{2}VAE}
    \label{fig:enter-label}
\end{figure*}
\subsection{Multi-View Generator}
In this subsection, we will illustrate the method to generate the common view and unique view feature representations, respectively. Given an item $v_i$, we define its' ID embedding $e_i \in \mathbb{R}^{d}$, a multi-hot vector $l_i \in \mathbb{R}^n$ representing categorical attribute feature and corresponding transformed image feature $c_i \in \mathbb{R}^{d}$. Meanwhile, we create an attribute embedding $A \in \mathbb{R}^{n \times d}$. Then, we encode the multi-typed item feature into latent variable as follows:
\begin{equation}
    \mu_{i_{e}} = e_{i}W_{\mu_{e}} + b_{\mu_e}, \quad \sigma_{i_{e}} = e_{i}W_{\sigma_{e}} + b_{\sigma_e}
\end{equation}
\begin{equation}
    \mu_{i_{c}} = c_{i}W_{\mu_{c}} + b_{\mu_c}, \quad \sigma_{i_{c}} = c_{i}W_{\sigma_{c}} + b_{\sigma_c}
\end{equation}
\begin{equation}
    a_i = AttenPooling(l_iA)
\end{equation}
\begin{equation}
    \mu_{i_{a}} = a_{i}W_{\mu_{a}} + b_{\mu_a}, \quad \sigma_{i_{a}} = a_{i}W_{\sigma_{a}} + b_{\sigma_a}
\end{equation}
Where $W_{\mu_{e}}, W_{\mu_{c}}, W_{\mu_{a}}, W_{\sigma_{e}}, W_{\sigma_{c}}, W_{\sigma_{a}}\in \mathbb{R}^{d\times d_{z}}$ and $b_{\mu_e}, b_{\mu_c}, b_{\mu_a}, b_{\sigma_e}, b_{\sigma_c}, b_{\sigma_a} \in \mathbb{R}^{d_z}$ are the parameters to learn. We use an attention-weighted sum to get the raw categorical attribute representation $a_i \in \mathbb{R}^d$.

After deriving the latent variables for both the categorical attribute feature and the item image feature, we employ the Product of Experts (PoE) to generate the common view across multi-typed features. As highlighted in Observation 1, the common view encapsulates the feature distribution that is shared and agreed upon by all feature types. PoE fusion extracts this common view by focusing on the overlapping high-probability regions of the individual distributions, effectively filtering out noise and inconsistencies. This capability makes PoE a robust tool for capturing the underlying common structure in heterogeneous data. Consequently, the latent variable of the common view, denoted as $z_{\text{com}} \sim \mathcal{N}(\mu_{\text{com}}, \sigma_{\text{com}}^2)$ is computed as follows:
We begin by concatenating $\mu_{i_a}, \mu_{i_c}$ and $\sigma_{i_a}, \sigma_{i_c}$ along their first axis, and then pass the concatenated result into a Product of Experts (PoE) fusion mechanism.
\begin{equation}
    T_{i_m} = \frac{1}{exp(\sigma_{i_m})+\epsilon}
\end{equation}
Where $T_{i_m}$ is the precision of the Gaussian expert in feature type $m\in \{a, c\}$ and $\epsilon$ is a small constant to avoid division by zero. Then, $\mu_{\text{com}}$ is calculated by the weighted average of the means, where the weights are the precisions, and $\sigma_{\text{com}}$ is calculated by the inverse of the sum of the precisions.
\begin{equation}
    \mu_{\text{com}} = \frac{\sum\limits_{m\in \{a,c\}}(\mu_{i_m} \cdot T_{i_m})}{\sum\limits_{m\in \{a,c\}}T_{i_m}}
    , \quad \sigma_{\text{com}} = log(\frac{1}{\sum\limits_{m\in \{a,c\}}T_{i_m}})
\end{equation}
Subsequently, we obtain the common view feature distribution $z_{\text{com}} \sim \mathcal{N}(\mu_{\text{com}}, \sigma_{\text{com}}^2)$ along with the unique view feature distributions for attribute feature $z_{a} \sim \mathcal{N}(\mu_{i_a}, \sigma_{i_a}^2)$ and image feature $z_{c} \sim \mathcal{N}(\mu_{i_c}, \sigma_{i_c}^2)$.
To enhance the disentanglement between the common view and unique views, we introduce a contrastive loss for the generated feature distributions. Specifically, we employ the reparameterization\cite{reparameter} trick to sample latent representations from a standard normal distribution $\epsilon \sim \mathcal{N}(0,I)$, rather than directly sampling from $\mathcal{N}(\mu_i, \sigma_i^2)$.
\begin{equation}
    z_{m} = \mu_{m} + \sigma_{m}\odot \epsilon ,\quad m\in\{a,c,\text{com}\}
\end{equation}
Subsequently, we detail the design of our proposed contrastive loss. In VAE-based methods, the goal is to accurately reconstruct the input data by minimizing the reconstruction loss within the ELBO framework. To this end, in our disentangled contrastive loss, the positive pairs of unique-view latent representations are defined as ($z_a, a_i$) and ($z_c, c_i$),respectively, while the negative pairs are ($z_a, z_{\text{com}}$) and ($z_c, z_{\text{com}}$). This design ensures that the latent variables effectively capture the essential information of the input data while promoting the decoupling of the latent representations of the common view and unique views. Specifically, the disentangled contrastive loss between the each unique view and the common view is computed as follows:
\begin{equation}
    \mathcal{L}_{v,\text{com}} = -log\frac{exp(sim(z_v,v_i)/\tau)}{exp(\sum_{j=1}^{B}{sim(z_v, z_{\text{com}_{j}}})/\tau)}, \quad v\in \{a,c\}
\end{equation}
Where $B$ represents the batch size, and the disentangled contrastive loss is calculated by sum of each unique and common view pair as:
\begin{equation}
    \mathcal{L}_{\text{disentangle}} = \sum\limits_{v\in \{a,c\}} \mathcal{L}_{v,\text{com}}
\end{equation}

\subsection{Multi-View Fusion}
After obtaining the multi-view latent representations, an effective fusion method is essential to derive a final item representation $z_{final} \sim \mathcal{N}(\mu_{final}, \sigma_{final}^2)$. Existing VAE-based models propose various fusion strategies. GoRec uses early fusion by concatenating heterogeneous features in the input space, while MVGAE applies a Product of Experts (PoE) in the latent space, distinguishing modalities by uncertainty levels. However, as noted in Observation 2, these approaches overlook diverse user preferences for different views and may lead to entanglement due to PoE’s mathematical properties. We now introduce our novel fusion method to address these issues.

We first perform a modality-level mixture-of-experts (M-MoE) to adaptively fuse the attribute and image views based on user preference. Given the target user embedding $u_t \in \mathbb{R}^d$, we apply modality-specific self-gating units to obtain filtered user representations:
\begin{equation}
    u_{v} = u_{t} \odot \sigma\big(u_{t} W_{g}^{v} + b_{g}^{v}\big), \quad v \in \{a, c\},
\end{equation}
where $W_{g}^{v} \in \mathbb{R}^{d \times d}$ and $b_{g}^{v} \in \mathbb{R}^{d}$ are learnable parameters. These gated embeddings capture user inclination toward each modality.

To compute the modality-level gating scores, we model the interaction among the user, the unique view, and the common view as $u_v \odot z_v \odot z_{\text{com}}$. The logits for each view are then obtained via linear projections:
\begin{equation}
    \text{logit}_{v} = \big( (u_{v} \odot z_{v} \odot z_{\text{com}}) w_{v} \big), \quad v \in \{a, c\},
\end{equation}
where $w_{v} \in \mathbb{R}^{d}$ is a learnable weight vector. Applying softmax over these logits yields the modality-level weights:
\begin{equation}
    [\text{gate}_a, \text{gate}_c] = \text{softmax}\big([\text{logit}_a, \text{logit}_c] / \tau\big),
\end{equation}
with temperature $\tau > 0$ controlling the sharpness of the distribution. The fused unique-view representation is then:
\begin{equation}
    x_{\text{unique}} = \text{gate}_a \cdot x_a + \text{gate}_c \cdot x_c,\quad x\in\{\mu,\sigma\}
\end{equation}

Subsequently, we perform a view-level MoE (V-MoE) that blends the common view $z_{\text{com}}$ and the aggregated unique view $z_{\text{unique}}$. A user-controlled scalar gate is computed as:
\begin{equation}
    \alpha = \sigma(\text{MLP}(u_t^\top v_{\text{com}})),
\end{equation}
where $v_{\text{com}} \in \mathbb{R}^{d}$ is a learnable parameter vector and $\sigma(\cdot)$ denotes the sigmoid function. The final representation is obtained through a convex combination:
\begin{equation}
    x_f = \alpha \cdot x_{\text{com}} + (1 - \alpha) \cdot x_{\text{unique}},\quad x\in\{\mu,\sigma\}.
\end{equation}
This two-stage hierarchical MoE, first over modalities within the unique views, then over the unique versus common views, enables fine-grained personalization while preserving shared semantic structure.

To obtain an approximated joint posteriors of $z\sim q_{\phi}(z|e, a, c)$, we also fuse the $z_e$ with $z_f$ by a simple MoE as follows,
\begin{equation}
    \mu=\mu_{i_e} + \mu_f, \quad \sigma=\sigma_{i_e} + \sigma_f
\end{equation}
\begin{equation}
    z=\mu + \sigma \odot \epsilon
\end{equation}
\subsubsection{Decoder}
After getting the approximated joint posteriors, we need to create a decoder $q_{\phi}(e|a,c,z)$ to reconstruct the item representation $e_{i_{new}} \in \mathbb{R}^d$. Under the framework of Conditional Variational AutoEncoder\cite{NIPS2015_8d55a249}, we define the decoder as follows:
\begin{equation}
    e_{i_{new}}=\mathcal{F}([z; a_i; c_i])
\end{equation}
Where $\mathcal{F}$ is a simple MultiLayer Perception (MLP).
\subsection{Co-occurrence Signal Injection}
In this section, our objective is to incorporate co-occurrence signals between users and warm items into the model. Existing VAE-based cold-start models, such as GoRec and CVAR, primarily rely on a pretraining module to learn co-occurrence representations from users and warm items, which significantly increases both computational time and parameter complexity. To address this limitation, we propose an end-to-end approach that injects co-occurrence signals directly into the generated item representation $e_{i_{new}}$ for new items using contrastive learning inspired by \cite{CLCRec,CCFCRec}. Positive items are defined as those with which the user has interacted, while negative items refer to those that have no interaction history with the target user. Specifically, for each positive item, we randomly select $c_p$ positive warm items and $c_n$ negative warm items from users' historical interactions $\mathcal{V}_u$. The contrastive loss $\mathcal{L}_{co}$ for co-occurrence signal injection is formulated as follows:
\begin{equation}
    pos = exp(sim(e_{i_{new}},e_{v^+})
\end{equation}
\begin{equation}
    neg = \sum\limits_{v^-\in \mathcal{V}_{u}^-} exp(sim(e_{i_{new}}, e_{v^-}))
\end{equation}
\begin{equation}
    \mathcal{L}_{\text{co}} = -\mathbb{E}_{v\in\mathcal{D}_{cold},v^+\in \mathcal{V}_u^+}[log\frac{pos}{pos+neg}]
\end{equation}
\subsection{Optimization}
In this section, we illustrate the training objective of our M\textsuperscript{2}VAE. M\textsuperscript{2}VAE follow the framework of Conditional Variational AutEncoder(CVAE), the formulation of our training objective is shown as:
\begin{align}
    \mathcal{L}_{\text{ELBO}}=-\mathbb{E}_{q_{\phi}(z|e,a,c)}[logp_{\theta}(e|a,c,z)] 
    \\
    +KL(q_{\phi}(z|e,a,c)||p_{\theta}(z_f|a,c))
     \\
    +\sum\limits_{v\in \{a,c\}}KL(q_{\phi}(z_v|v)||p_{\theta}(z_v))
\end{align}
Where the reconstruction loss $-\mathbb{E}_{q_{\phi}(z|e,a,c)}[logp_{\theta}(e|a,c,z)]$ is defined as follows:
\begin{equation}
    -\mathbb{E}_{q_{\phi}(z|e,a,c)}[logp_{\theta}(e|a,c,z)]=\text{MSE}(e_i,e_{i_{new}})
\end{equation}
To address the challenges of cold-start item recommendation, we employ the Bayesian Personalized Ranking (BPR) loss for optimizing our new item representation. The BPR loss is defined as follows:
\begin{equation}
    \mathcal{L}_{\text{BPR}} = \sum\limits_{(i,u,u^-)\in \mathcal{D}}-log(e_{i_{new}}e_u^T-e_{i_{new}}e_{u^-}^T)
\end{equation}
Thus, our final training objective is calculated as:
\begin{equation}
    \mathcal{L}=\mathcal{L}_{\text{ELBO}} + \mathcal{L}_{\text{BPR}}+\alpha \mathcal{L}_{\text{disentangle}} + \beta \mathcal{L}_{\text{co}}
\end{equation}
where $\alpha$ and $\beta$ are hyperparameters.
\begin{figure}
    \centering
    \includegraphics[width=1.05\linewidth]{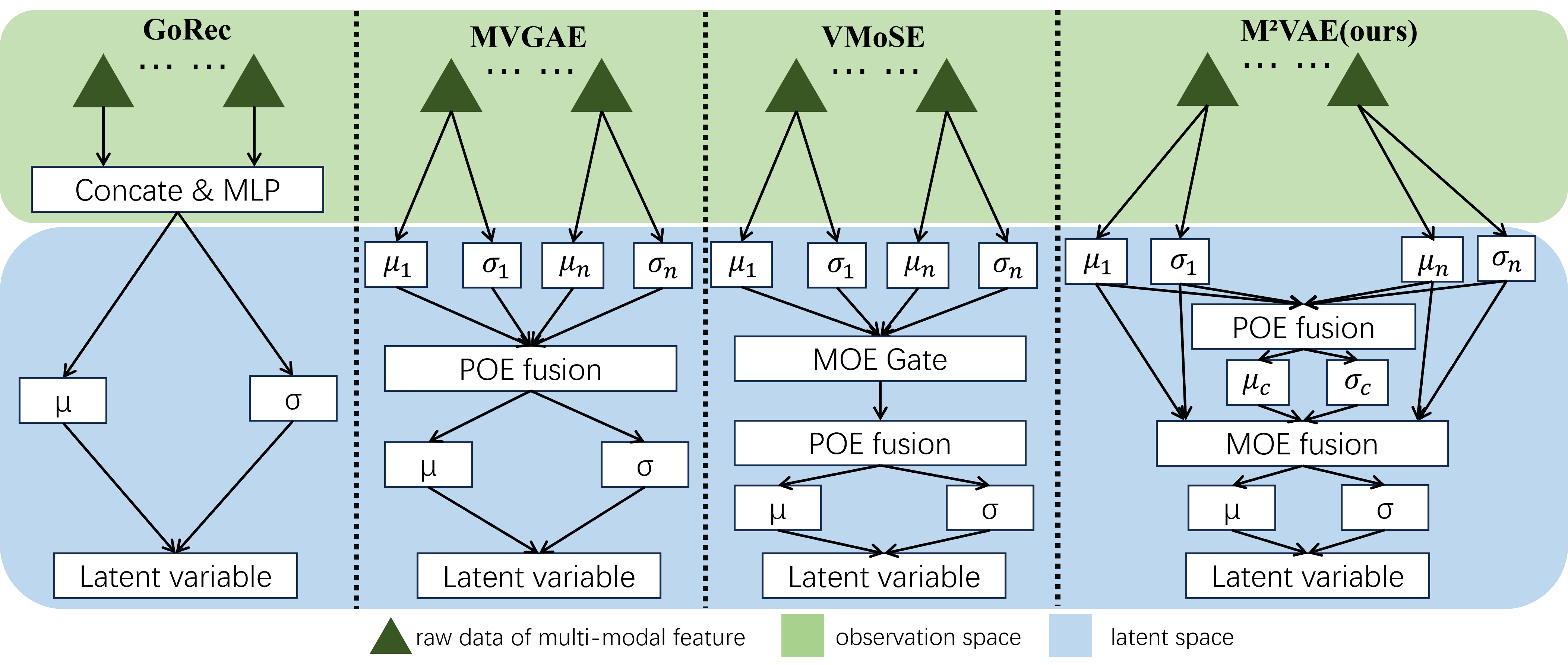}
    \caption{Structural comparison on multi-modal fusion with existing vae-based methods}
    \label{fig:architechture_comparision}
\end{figure}
\begin{table}
\scriptsize
\centering
\begin{tabular}{l|c|c|c|c}
\hline
\textbf{Property} & \textbf{GoRec} & \textbf{MVGAE} & \textbf{VMoSE} & \textbf{M²VAE} \\ 
\hline
Modality Uncertainty Modeling & \XSolidBrush & \checkmark & \checkmark & \checkmark \\
Dynamic Weighting Mechanism & \XSolidBrush & \XSolidBrush & \checkmark & \checkmark \\
Common-Specific Disentanglement & \XSolidBrush & \XSolidBrush & \XSolidBrush & \checkmark \\
\hline
\end{tabular}
\label{tab:comparison_summary}
\caption{Key Characteristics of Fusion Architectures}
\end{table}
\begin{table*}
    \centering
    \begin{tabular}{c|c|c|c}
    \hline
        Process &  M\textsuperscript{2}VAE & MVGDAE & GoRec \\
        \hline
        Train &  $O(Md^2+Md+Hd+b^{2}d)$ & $O(MLNd^{2}+Md+Hd)$ & $O(Hd)+O(pretrain)$\\
        \hline
        Inference & $O(Md^2+Md+Hd)$ & $O(MLNd^{2}+Md+Hd)$ & $O(knd+Hd)$\\
        \hline
    \end{tabular}
    \caption{Time complexity comparison }
    \label{tab:time_complexity}
\end{table*}

\subsection{Discussion}
\subsubsection{Fusion Architecture}
We highlight the fusion architectures of existing methods besides our M\textsuperscript{2}VAE in Fig. 3 and structural advantages of M\textsuperscript{2}VAE's fusion mechanism over key baselines in Table 1. A structural comparison reveals critical limitations in prior works.

Specifically, GoRec's simple feature concatenation is deterministic, entirely ignoring modality-specific uncertainty. While MVGAE's Product-of-Experts (PoE) addresses uncertainty, it treats all modalities equally and conflates common and specific features into a single consensus, rendering it vulnerable to noisy inputs. VMoSE advances this by incorporating a gating network for dynamic weighting, yet it still performs a "flat" fusion, failing to explicitly disentangle the shared semantics from unique characteristics.

M²VAE is designed to systematically overcome these limitations through a novel hierarchical architecture. It first leverages PoE to isolate a robust common view, the semantic consensus across modalities. Subsequently, a user-aware hierarchical Mixture-of-Experts (MoE) adaptively fuses this shared knowledge with informative specific views. This two-stage process ensures that the final representation is not only robust and flexible but also inherently more interpretable, as summarized in Table 1.


\subsubsection{Time Complexity}
The overall inference and training time complexity comparison with existing methods are illustrated in the Table 2. Specifically, the time complexity of MoE, PoE, VAE, Contrastive Loss, Kmeans and GCN are $O(d^2)$, $O(d)$, $O(Hd)$, $O(b^{2}d)$, $O(knd)$ and $O(LND^2)$. $M$ is the number of modality, $H$ is hidden layer size, $b$ is batch size, $d$ is feature dimension, $k$ is number of clusters and $n$ is number of items. M\textsuperscript{2}VAE’s inference complexity avoids the expensive GCN propagation required by MVGDAE. For training, M\textsuperscript{2}VAE introduces a contrastive loss overhead but avoids the pretraining dependencies of GoRec. In general, M\textsuperscript{2}VAE demonstrates comparable efficiency to SOTA methods.

For scalability, MoE takes $Md^2$ of parameters number, which is comparable to the GCN encoder($O(LND^2)$) used in MVGDAE. Moreover, the pretrain module in GoRec takes extra parameters. In general, our proposed M\textsuperscript{2}VAE is also comparable to SOTA methods in space complexity.


%% file: section/experiment.tex
\section{Experiment}
In this section, we conduct comprehensive experiments on three real-world datasets to answer the following questions:
\begin{itemize}
    \item RQ1: How does our proposed M\textsuperscript{2}VAE perform compared with other
     state-of-the-art baselines on multi-modal cold-start recommendation scenarios?
    \item RQ2: How do different designed components play roles in our proposed model?
    \item RQ3: What interpretability insights can we uncover from case study?
\end{itemize}

\subsection{Experimental Settings}
\begin{table*}[htbp]
  \scriptsize
  \label{tab:freq}
  \begin{tabular}{c|cccc|cccc|cccc}
    \toprule
    \multirow{2}{*}{Model} & \multicolumn{4}{c|}{ML-20M} & \multicolumn{4}{c}{Video\&Games} & \multicolumn{4}{|c}{Sports\&Outdoors} \\
    \cline{2-5} \cline{6-9} \cline{10-13}
    \quad & HR@5 & NDGC@5 & HR@10 & NDGC@10  & HR@5 & NDGC@5 & HR@10 & NDGC@10  & HR@5 & NDGC@5 & HR@10 & NDGC@10  \\
    \midrule 
    \multicolumn{13}{l}{\textit{Contrastive Learning-based Method}} \\   \hline
    CLCRec & 0.2677& 0.4695 & 0.2371 & 0.4820 & 0.0229 & 0.0646 & 0.0189 & 0.0769 & 0.0098 & 0.0289 &0.0074 &0.0368 \\
    CCFCRec & \underline{0.2969} & \underline{0.4798} & \underline{0.2592} & \underline{0.4933} & \underline{0.0326} & \underline{0.0916} & \underline{0.0260} & \underline{0.1074} & 0.0107 & 0.0310 & 0.0085 & 0.0386\\
    PAD-CLRec & 0.2952 & 0.4699 & 0.2501 & 0.4896 & 0.0313 & 0.0889 & 0.0247 & 0.1005 & 0.0101 & 0.0297 & 0.0082 & 0.0364 \\
    \hline
    \multicolumn{13}{l}{\textit{General Generative Method}} \\   \hline
    DropoutNet & 0.2264 & 0.3836 & 0.1993 & 0.4065 & 0.0153 & 0.0417 & 0.0099 & 0.0453 & 0.0057 & 0.0153 & 0.0044 & 0.0189 \\
    MTPR & 0.2701 & 0.4504 & 0.2393 & 0.4588 & 0.0161 & 0.0457 & 0.0112 & 0.0518 & 0.0099 & 0.0292 & 0.0077 & 0.0370 \\
    \hline
    \multicolumn{13}{l}{\textit{GAN-based Method}} \\   \hline
    LARA & 0.2425 & 0.4595 & 0.2165 & 0.4541 & 0.0140 & 0.0370 & 0.0074 & 0.0381 & 0.0069 & 0.0162 & 0.0055 & 0.0193 \\
    CVAR & 0.2363 & 0.4301 & 0.2188 & 0.4359 & 0.0149 & 0.0405 & 0.0109 & 0.0481 & 0.0090 & 0.0347 & 0.0073 &0.0569 \\
    GAR & 0.2207 & 0.4216 & 0.2053 & 0.4229 & 0.0131 & 0.0396 & 0.0101 & 0.0441 & 0.0082 & 0.0343 & 0.0064 & 0.0575 \\
    \hline
    \multicolumn{13}{l}{\textit{VAE-based Method}} \\   \hline
    CVAE & 0.2306 & 0.4285 & 0.2147 & 0.4309 & 0.0143 & 0.0399 & 0.0102 & 0.0480 & 0.0059 & 0.0151 &0.0043  & 0.0188 \\
    MVDGAE & 0.2789 & 0.4586 & 0.2453 & 0.4660 & 0.0161 & 0.0456 & 0.0126 & 0.0539 & 0.0105 & 0.0313 & 0.0088 & 0.0390\\
    MVGAE & 0.2756 & 0.4561 & 0.2433 & 0.4606 & 0.0239 & 0.0712 & 0.0193 & 0.0781 & 0.0128 & 0.0477 & 0.0083 & 0.0609\\
    VMoSE & 0.2912 & 0.4720 & 0.2503 & 0.4891 & 0.0299 & 0.0874 & 0.0219 & 0.0996 & 0.0147 & 0.0505 & 0.0100 & 0.0628\\
    GoRec & 0.2908 & 0.4715 & 0.2493 & 0.4873 & 0.0301 & 0.0877 & 0.0228 & 0.1012 & \underline{0.0166} & \underline{0.0529} & \underline{0.0123} & \underline{0.0655}\\
    \hline
    M\textsuperscript{2}VAE & \textbf{0.3045} & \textbf{0.4805} & \textbf{0.2656} & \textbf{0.5031} & \textbf{0.0367} & \textbf{0.1009} & \textbf{0.0286} & \textbf{0.1161} & \textbf{0.0192} & \textbf{0.0561} & \textbf{0.0157} & \textbf{0.0689} \\
    \hline
    Impr. & 2.6\% & 1.5\% & 2.5\% & 2.0\% & 12.6\% & 10.2\% & 10.0\% & 8.1\% & 15.7\% &6.0\% & 27.6\% & 5.2\% \\
  \bottomrule
\end{tabular}
  \caption{Experimental results on three datasets. The best results are boldfaced and the second-best results are underlined}
\end{table*}

\subsubsection{Dataset}
To evaluate the effectiveness of M\textsuperscript{2}VAE, we conduct experiments on three real-world datasets: Movielens-20M, Amazon Video\&Game, and Amazon Sports\&Outdoors. These datasets vary in size and sparsity, providing a comprehensive testbed for our model.
\subsubsection{Hyperparameter Settings}
We implement M\textsuperscript{2}VAE and all baselines in PyTorch. For models requiring pretraining, we use a 64-dimensional preference representation, batch size of 1024, and a 2-layer MLP. In M\textsuperscript{2}VAE, the latent dimension $d$ is 128. Hyperparameters $\alpha$ and $\beta$ are tuned via grid search, yielding $(10, 0.5)$ for ML-20M, $(100, 1)$ for Video\&Games, and $(100, 0.5)$ for Sports\&Outdoors. Training uses sampled positive/negative pairs: (10, 40) for ML-20M, (5, 40) for Video\&Games, and (5, 20) for Sports\&Outdoors. All models are fairly tuned and evaluated over 5 runs with averaged results.
\subsubsection{Baselines}
The baseline models can be divided into several categories: (1) Contrastive learning: CLCRec, CCFCRec, PAD-CLRec. (2) General generative: DropoutNet, MTPR. (3) GAN-based: LARA, CVAR, GAR. (4) VAE-based: MVDGAE, GoRec, CVAE, MVGAE, VMoSE.
\subsection{Performance Comparison(RQ1)}
The experimental results demonstrate that M\textsuperscript{2}VAE significantly outperforms baseline methods across ML-20M, Video\&Games, and Sports\&Outdoors datasets in cold-start recommendation tasks. Our model achieves superior performance in HR@5/10 and NDCG@5/10 metrics, with relative improvements over the second-best models: 2.6\%/2.5\% and 1.5\%/2.0\% (ML-20M), 12.6\%/10.0\% and 10.2\%/8.1\% (Video\&Games), and up to 15.7\%/27.6\% and 6.0\%/5.2\% (Sports\&Outdoors). The effectiveness stems from its disentangled representation learning and adaptive fusion mechanisms. Moreover, we make extra observations and analyses as following: 
\begin{itemize}
    \item VAE-based methods generally outperform GAN-based approaches in cold-start recommendation tasks. This stems from their probabilistic framework that explicitly models data distributions, effectively handling uncertainty and sparse/incomplete data—key challenges in cold-start scenarios. Conversely, GANs lack explicit distribution modeling, limiting their ability to capture underlying patterns from limited interactions.
    \item We observe that contrastive learning (e.g., CCFCRec) captures co-occurrence signals more effectively than pretraining (GoRec, CVAR) in data-scarce scenarios. On ML-20M (35k interactions) and Video\&Games (24k), CCFCRec achieves superior performance. However, GoRec excels on Sports\&Outdoors with 237k interactions. This suggests that contrastive learning better leverages limited data for pattern extraction, while pretraining requires larger interaction volumes to achieve comparable results.
\end{itemize}
\begin{table}
  \setlength{\tabcolsep}{2pt}
  \centering
  \small
  \label{tab:freq}
  \begin{tabular}{c|cc|cc|cc}
    \toprule
    \multirow{2}{*}{Model Variants} & \multicolumn{2}{c|}{ML-20M} & \multicolumn{2}{c}{Video\&Games} & \multicolumn{2}{c}{Sports\&Outdoors}\\
    \cline{2-3} \cline{4-5} \cline{6-7}
    \quad & H@5 & N@5 & H@5 & N@5 & HR@5 & N@5\\
    \midrule
    M\textsuperscript{2}VAE & 0.3045 & 0.4805 & 0.0367 & 0.1009 & 0.0192 & 0.0561\\
    \hline
    \textit{w/o} \textbf{common} & 0.2673 & 0.4373 & 0.0297 & 0.0841 & 0.0164 & 0.0493\\
    \textbf{early generate} & 0.2798 & 0.4593 & 0.0303 & 0.0899 & 0.0179 & 0.0520\\
    \hline
    \textit{w/o} \textbf{SG} & 0.2801 & 0.4629 & 0.0308 & 0.0927 & 0.0185 & 0.0546\\
    \textit{w/o} \textbf{M-MoE} & 0.2827 & 0.4633 & 0.0312 & 0.0929 & 0.0188 & 0.0549\\
    \textit{w/o} \textbf{V-MoE} & 0.2845 & 0.4650 & 0.0313 & 0.0931 & 0.0192 & 0.0550\\
    \textbf{weighted PoE}& 0.2650 & 0.4393 & 0.0294 & 0.0871 &  0.0160 & 0.0491\\
    \hline
    \textit{w/o} \textbf{DCL} & 0.2783 & 0.4445 & 0.0303 & 0.0855 & 0.0172 & 0.0499\\
    \hline
    \textit{w/o} \textbf{Co-CL} & 0.2756 & 0.4390 & 0.0302 & 0.0847 & 0.0167 & 0.0494\\
  \bottomrule
\end{tabular}
\caption{Ablation study with key modules}
\end{table}
\begin{figure}
  \centering
  \subfigure[\textit{w/o} DCL/Sports]{
  \includegraphics[scale=0.13]{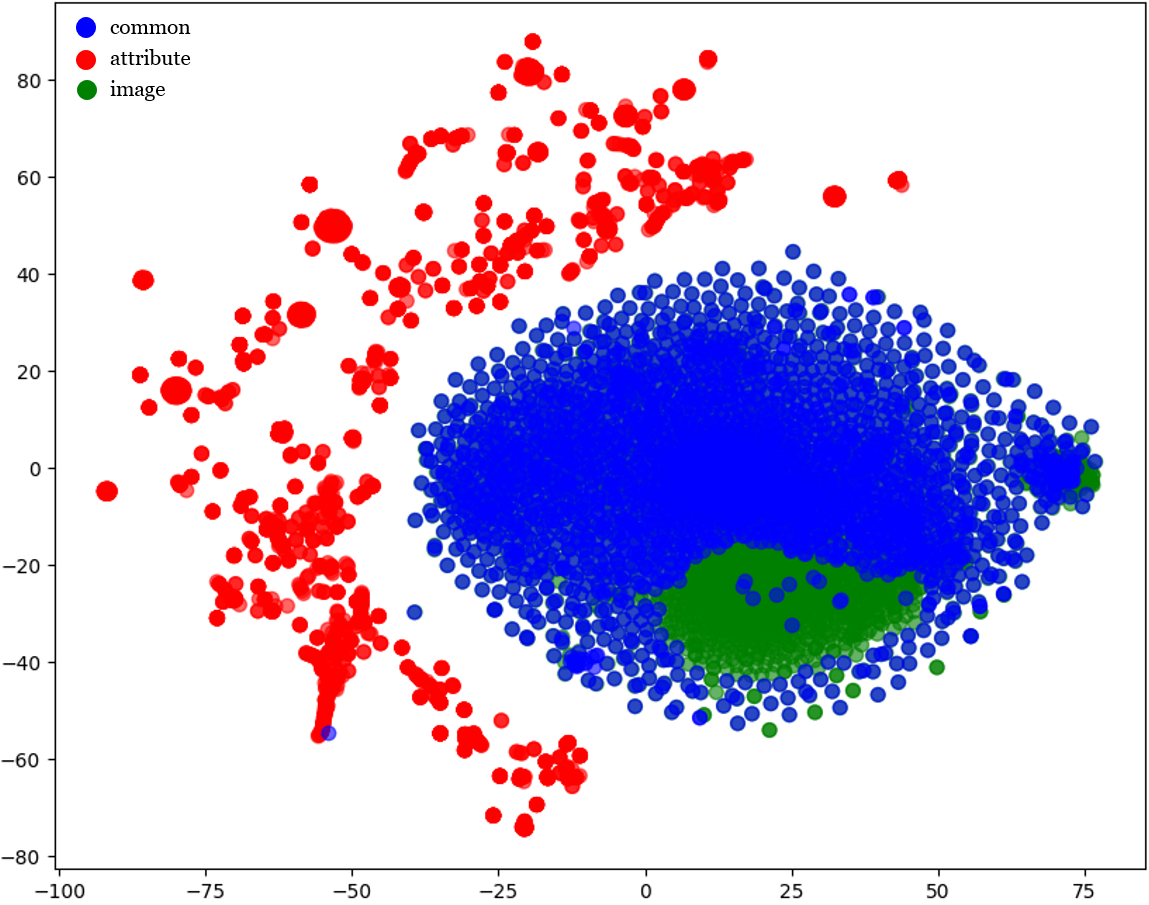}
  }
  \subfigure[\textit{w/o} DCL/Video]{
  \includegraphics[scale=0.13]{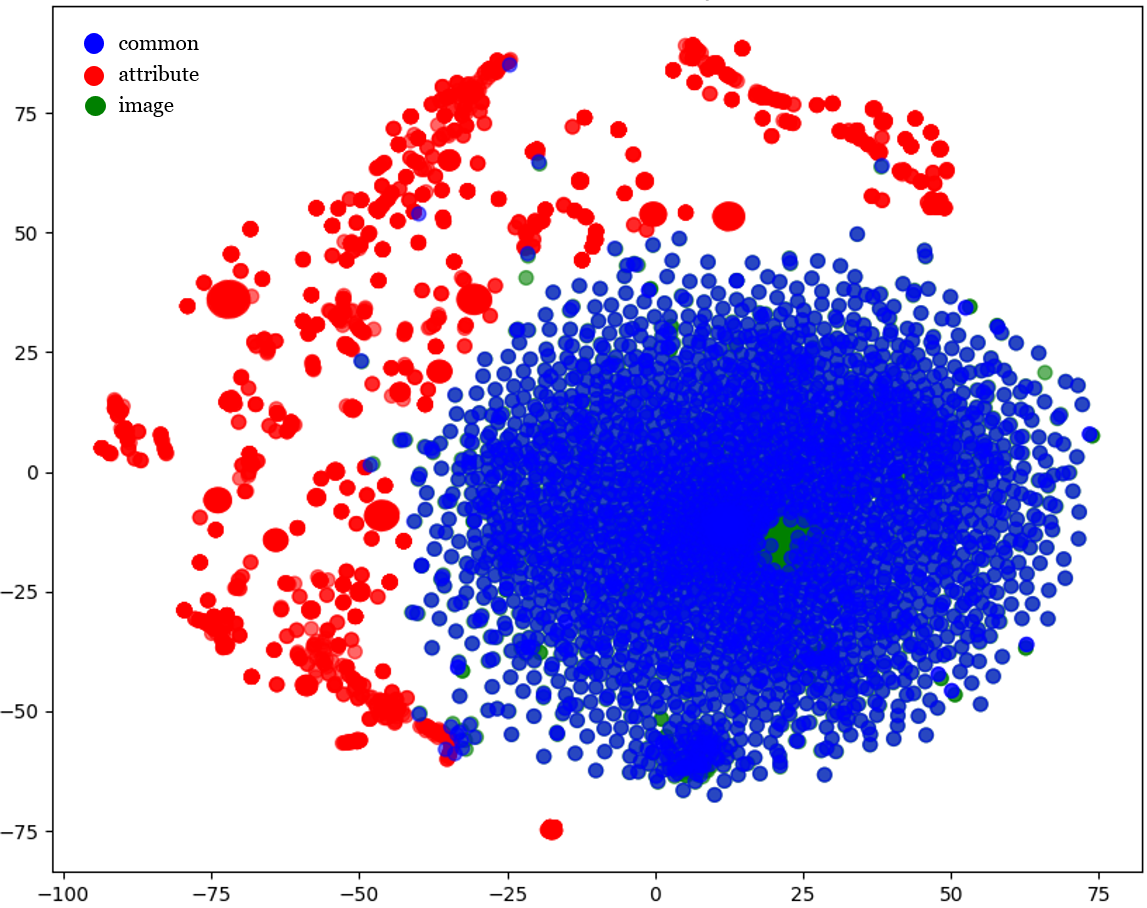}
  }\
  \subfigure[\textit{w/o} DCL/ML]{
  \includegraphics[scale=0.13]{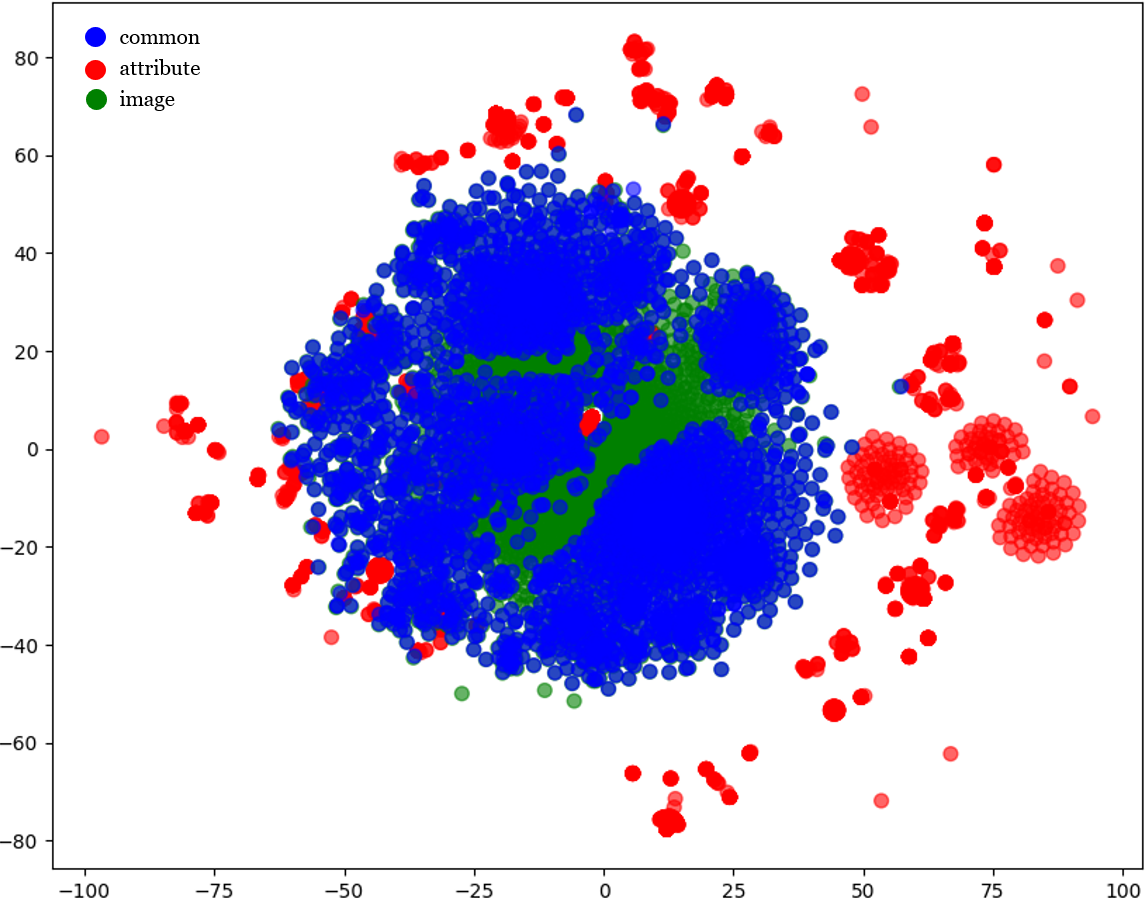}
  }
  \subfigure[with DCL/Sports]{
  \includegraphics[scale=0.13]{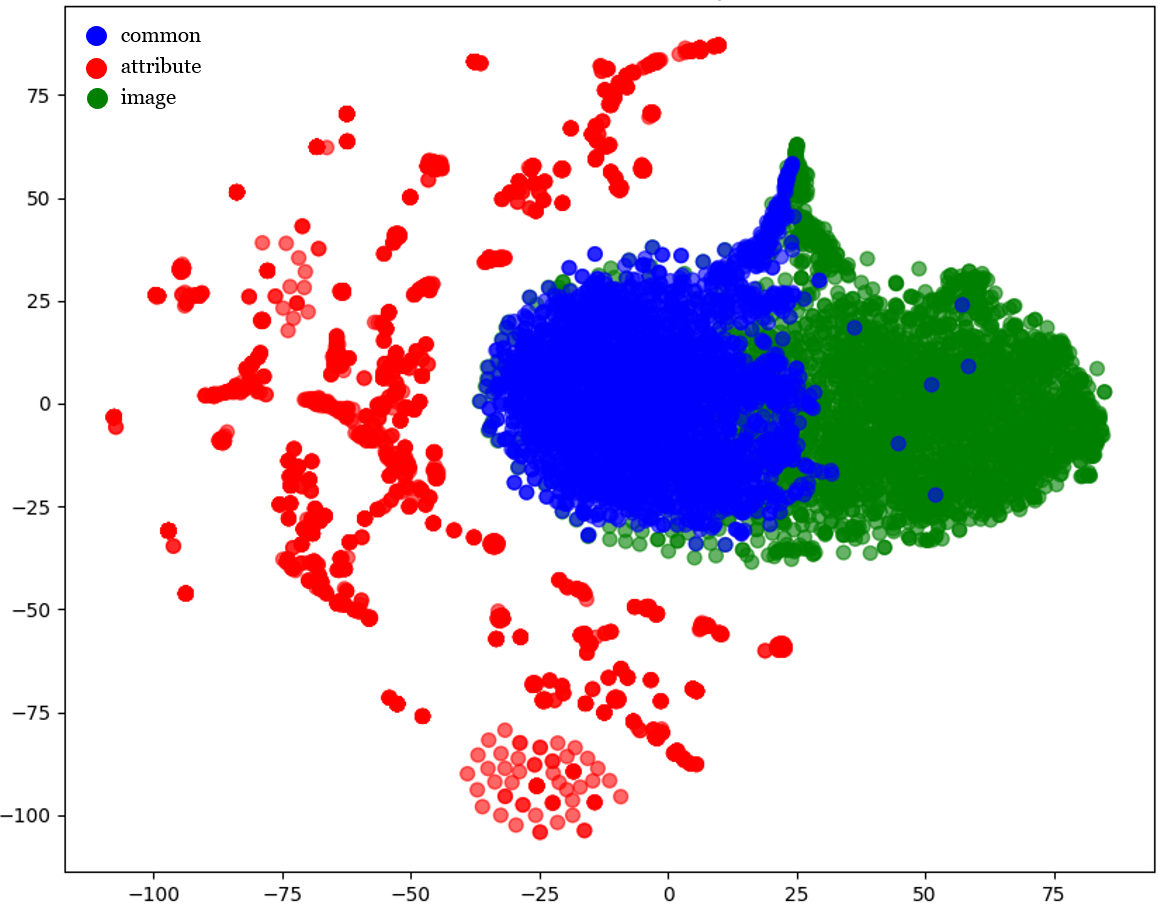}
  }
  \subfigure[with DCL/Video]{
  \includegraphics[scale=0.13]{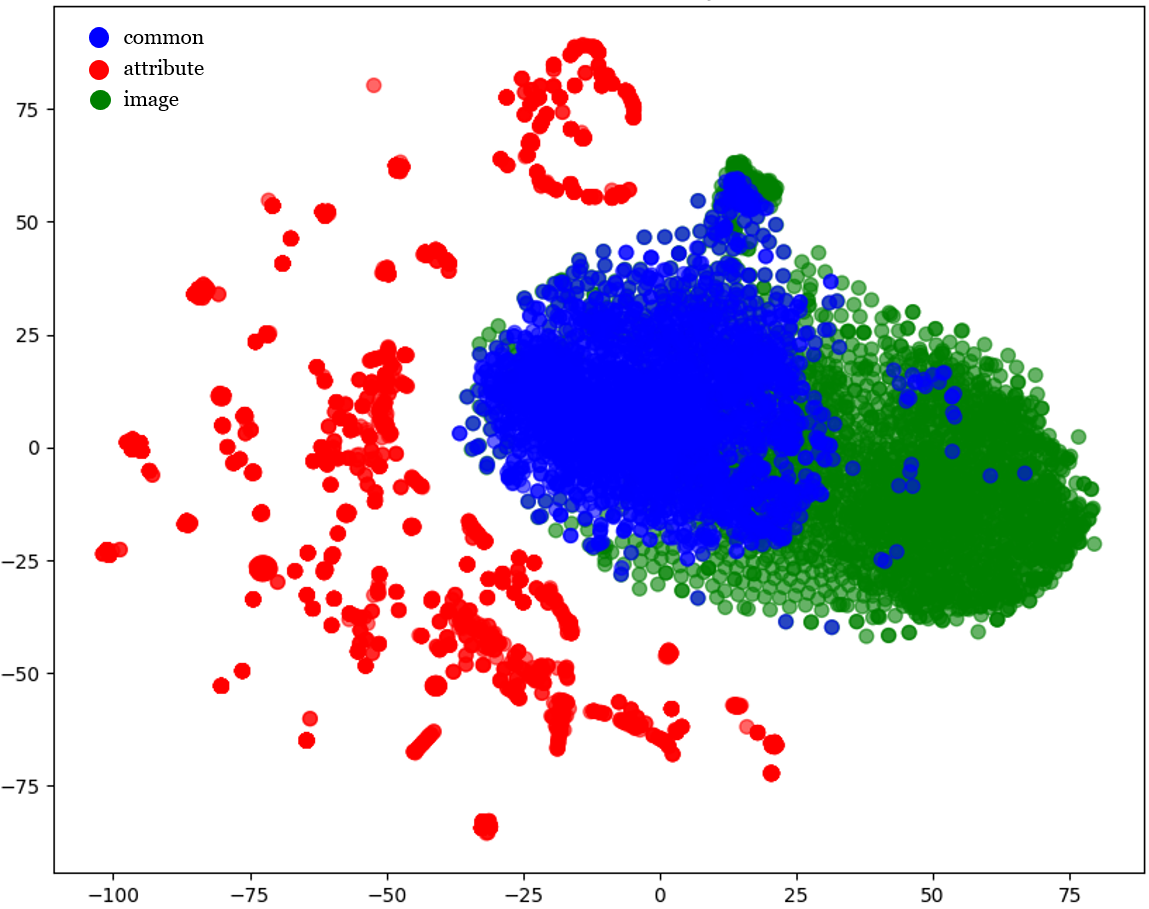}
  }
  \subfigure[with DCL/ML]{
  \includegraphics[scale=0.13]{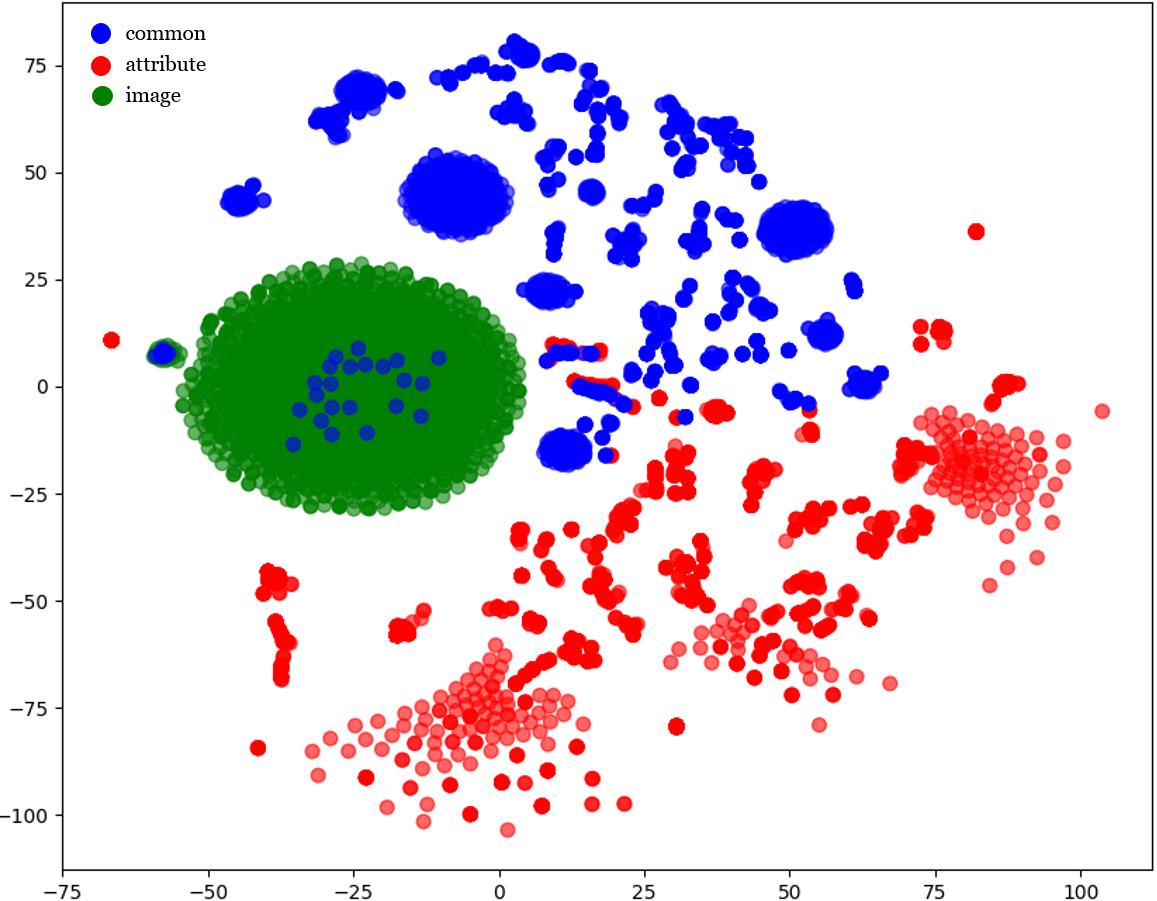}
  }
  \caption{T-SNE of multi-view representations \textit{w/o} DCL}
  \vspace{-0.4cm}
\end{figure}
\subsection{Ablation Study(RQ2)}
In this section, we show the ablation study of each key component.
\subsubsection{Effect Of Multi-View Generator Structure}
We evaluate our multi-view generator's impact on recommendation performance by comparing M\textsuperscript{2}VAE with two variants: one without common view (w/o common) and another using raw common view generation (early generate) following GoRec's approach. Results show that both common view generation and latent space learning are critical for optimal performance (Table 4). This validates the effectiveness of our PoE-based multi-view architecture in enhancing recommendation accuracy.
\subsubsection{Effect Of Multi-view Fusion Structure}
In this subsection, we examine the impact of our proposed multi-view fusion structure, which employs user-aware hierarchical Mixture of Experts (MoE) fusion to integrate common view and unique view representations. To validate the effectiveness of our fusion method, we introduce model variants including "\textit{w/o} SG", "\textit{w/o} M-MoE" and "\textit{w/o} V-MoE" to verify the effectiveness of each components. Additionally, we replace our MoE fusion with a weighted Product of Experts (PoE) fusion, similar to MVGAE as illustrated in Fig. 3, to further demonstrate the advantages of MoE in effectively fusing multi-view feature representations. These comparisons highlight the superior performance and flexibility of our proposed fusion approach.

\subsubsection{Effect Of Disentangled Contrastive Loss}
We evaluate the effectiveness of our disentangled contrastive loss (DCL) by comparing with a variant (w/o DCL). Results show significant performance degradation without DCL, with T-SNE visualizations revealing strong coupling between attribute/image features and common representations in the w/o DCL model (Fig. 4). While DCL achieves clearer disentanglement in ML-20M, residual overlaps persist in Sports\&Outdoors and Video\&Games. Notably, attribute features dominate common representations in ML-20M, whereas image features drive common space distribution in other datasets, demonstrating DCL's varied effectiveness across domains

\subsubsection{Effect Of Co-occurrence Contrastive Loss}
In this section, we examine the impact of the co-occurrence contrastive loss (Co-CL) on model performance. By removing the co-occurrence signal injection, we observe a significant decline in model performance. This result underscores the critical role that co-occurrence signals from warm items play in effectively recommending new items to users. The findings highlight the importance of incorporating such signals to enhance recommendation accuracy and robustness. 
\begin{figure}
    \centering
    \includegraphics[width=0.9\linewidth]{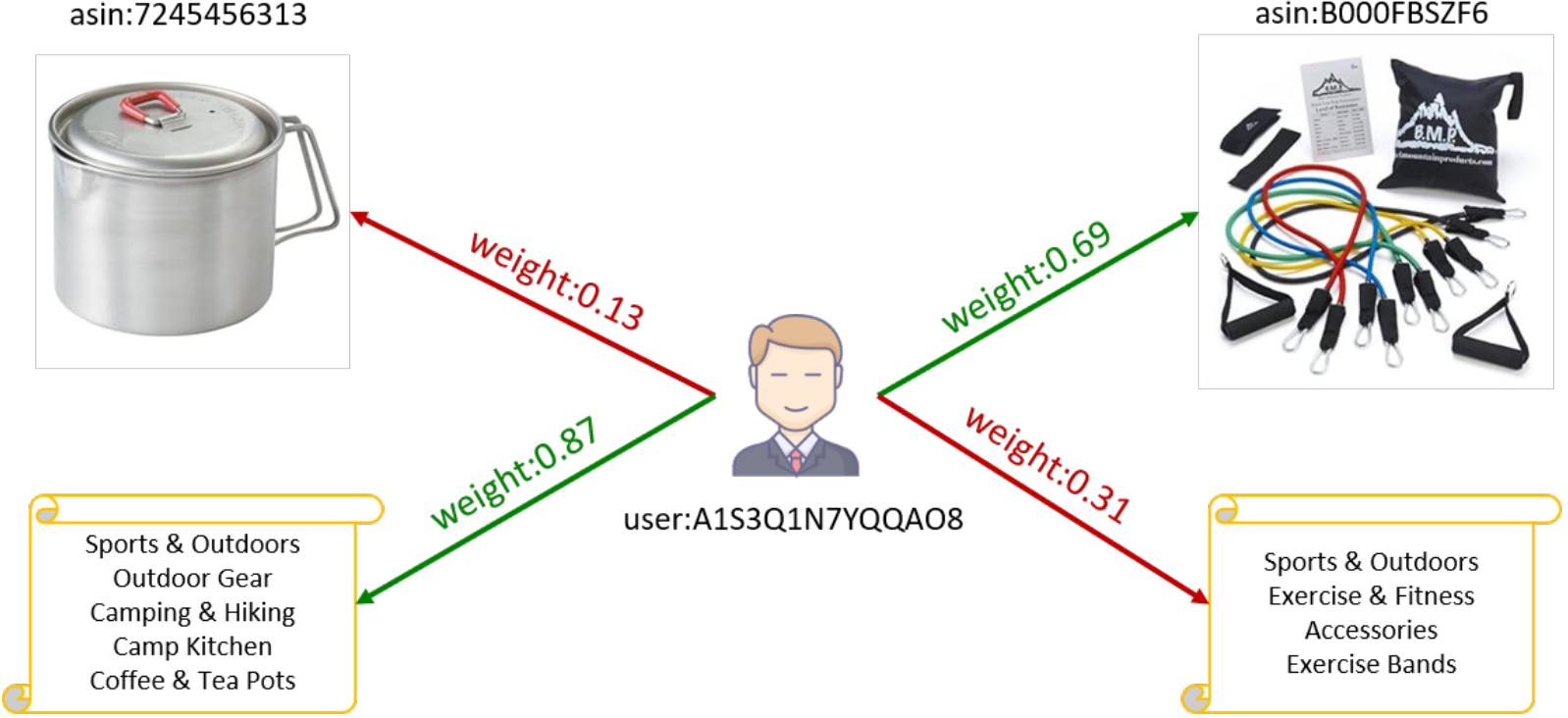}
    \caption{Case study of personalized inclination on Sports\&Games}
    \label{fig:enter-label}
    \vspace{-0.4cm}
\end{figure}
\subsection{Case Study(RQ3)}
In this subsection, we conduct a case study on the Sports\&Games dataset to explore interpretability insights. As illustrated in Fig. 5, we visualize the personalized inclination of user "A1S3Q1N7YQQAO8" toward two distinct unique views: categorical attributes and image features, modeled by the MoE gate in Eq. (12). Our analysis reveals that the user exhibits a stronger preference for the categorical attribute feature of item "7245456313," while the image feature of item "B000FBSZF6" captures significantly more attention compared to its attribute feature. We provide the following explanations for these observations:
(1) The categorical attribute feature of item "7245456313" contains richer and more valuable information, such as "Camp\&Kitchen" and "Coffee\&Tea Pots," which cannot be fully conveyed through its image feature. (2) Conversely, the image feature of item "B000FBSZF6" provides additional visual details, such as the color of bands and bags, that are not explicitly represented in its categorical attributes. This case study highlights the importance of modeling personalized inclinations toward different feature types, as it enables a more nuanced understanding of user preferences and enhances the interpretability of recommendation systems.

%% file: section/conclusion.tex
\section{Conclusion}
In this paper, we proposed the Multi-Modal Multi-View Variational AutoEncoder (M\textsuperscript{2}VAE), a novel generative model designed to enhance the representation of multi-typed item features in recommendation systems, which explicitly models both common and unique views of item features and incorporates personalized user inclinations through a Mixture of Experts (MoE) fusion mechanism. Our extensive experiments on benchmark datasets validate the effectiveness of M\textsuperscript{2}VAE in capturing complex feature interactions and improving recommendation accuracy.

%% file: section/ack.tex
\section*{Acknowledgments}
This work was supported by “Pioneer” R\&D Program of Zhejiang under Grant No.2023C01029, the National Key Research and Development Plan of China (2023YFB4502305) and Ant Group through Ant Research Intern Program.